\documentclass[conference]{IEEEtran}

\usepackage{cite}

\def\equ #1{\begin{equation}#1\end{equation}}
\def\spl #1{\begin{split}#1\end{split}}
\def\bma #1{\begin{bmatrix}#1\end{bmatrix}}
\def\cas #1{\begin{cases}#1\end{cases}}
\def\norm #1{\left\|#1\right\|}

\def\bgo{\bm{\Gamma_{\Omega}}}
\def\bOmega{\bm{\Omega}}
\def\bT{\mathbb{T}}
\def\bC{\mathbb{C}}

   \usepackage[dvips]{graphicx}

\usepackage[cmex10]{amsmath}
\usepackage{amssymb}

\newtheorem{theorem}{Theorem}
\newtheorem{remark}{Remark}

\usepackage{flushend}
\usepackage{color}
\usepackage{algorithm}
\usepackage{algorithmic}
\usepackage{subfigure}

\usepackage{array}

\usepackage{bm}
\usepackage{amsfonts}

\usepackage[font=footnotesize]{subfig}

\begin{document}

\title{Joint Doppler and Channel Estimation with Nested Arrays for Millimeter Wave Communications}

\author{\IEEEauthorblockN{Xiaohuan Wu\IEEEauthorrefmark{1}, Wei-Ping Zhu\IEEEauthorrefmark{1}\IEEEauthorrefmark{2}, Min Lin\IEEEauthorrefmark{1} and Jun Yan\IEEEauthorrefmark{1}}
\IEEEauthorblockA{\IEEEauthorrefmark{1} Key Laboratory of Ministry of Education in  Broadband Wireless Communication and Sensor Network Technology, \\
Nanjing University of Posts and Telecommunications, Nanjing, China}
\IEEEauthorblockA{\IEEEauthorrefmark{2}  Department of Electrical and Computer Engineering, Concordia University, Montreal, Canada}
}

\maketitle

\begin{abstract}
  Channel estimation is essential for precoding/combining in millimeter wave (mmWave) communications. However, accurate estimation is usually difficult because the receiver can only observe the low-dimensional projection of the received signals due to the hybrid architecture. We take the high speed scenario into consideration where the Doppler effect caused by fast-moving users can seriously deteriorate the channel estimation accuracy. In this paper, we propose to incorporate the nested array into analog array architecture by using RF switch networks with an objective of reducing the complexity and power consumption of the system. Based on the covariance fitting criterion, a joint Doppler and channel estimation method is proposed without need of discretizing the angle space, and thus the model mismatch effect can be totally eliminated. We also present an algorithmic implementation by solving the dual problem of the original one in order to reduce the computational complexity. Numerical simulations are provided to demonstrate the effectiveness and superiority of our proposed method.
\end{abstract}

\begin{IEEEkeywords}
Channel estimation, millimeter wave (mmWave), nested arrays, covariance fitting criterion.
\end{IEEEkeywords}

\IEEEpeerreviewmaketitle

\section{Introduction}
\label{sec Introduction}

The millimeter-wave (mmWave) enabled massive multiple-input multiple-output (MIMO) system is considered as a promising technique for future 5G wireless communications due to its provision of sufficient array gain for reliable communications \cite{WhatWill5GBe2014JSAC, Han2015ComMag, LinMin2018JSAC}. Due to the unaffordable hardware cost and limited space, however, the number of radio-frequency (RF) chains is seriously restricted. On the contrary, the number of antennas has to become huge in order to keep large antenna aperture. The large gap between the number of RF chains and that of antennas requires new precoding/combining strategy. The hybrid analog-digital architecture was proposed to solve this problem \cite{Molisch2017ComMag, Xiao2017JSAC}. The key idea of hybrid architectures is to divide the conventional digital precoding/combining with large size into two parts: a large-size analog precoding/combining and a dimension-reduced digital precoding/combining. The hybrid MIMO architectures proposed for mmWave communications are usually based on a network composed of a large number of phase shifters \cite{Heath2017JSAC} (full-connected network \cite{hybrid2014TWC_Heath} or sub-connected network \cite{Dai2016JSAC_subarray_connected}). However, the phase shifter-based network is not a simple circuit at mmWave \cite{Heath2016Access} and might not be able to adapt to the quick variation of instantaneous channels over time \cite{Molisch2017ComMag}.

An alternative MIMO architecture is the switch network \cite{Heath2016Access}, which was proposed to reduce the complexity and power consumption of the phase shifter network. It was shown in \cite{Heath2016Access} that the combining matrices based on switches provide equal or even lower coherence than the matrices associated to a phase shifter based architecture.

The hybrid precoding/combining design requires the acquisition of accurate channel state information (CSI), which, however, is not an easy task for the hybrid architecture, either with phase shifters or switches. This is because with the hybrid architecture, the receiver can only observe the low-dimensional representation of the signals received on the entire antennas, making the channel estimation for mmWave systems more difficult. To solve this problem, the sparse channel estimation methods exploiting the sparsity of mmWave MIMO channels have been proposed.
In particular, due to the poor scattering property of mmWave, the number of channel paths becomes very small. Hence the parametric physical channel model could be exploited to determine the angle-of-arrivals (AoAs) (a.k.a. direction-of-arrival (DOA) in the array signal processing) and the path gains. Then the CSI can be determined by the estimated AoAs and path gains and moreover, determining the AoAs can usually be transformed to a sparse recovery problem, resulting in sparse channel estimation methods \cite{Heath2014JSTSP, Heath2016Access}.
However, the sparse channel estimation method such as orthogonal matching pursuit (OMP) \cite{Heath2014JSTSP} requires to discretize the whole angle space with a pre-defined grid set, which may bring in high computations when the grid size is small. On the other hand, the switching network can be regarded as "antenna selection" process. From the sparse array theory, there exist some special structures in antenna selection which can be utilized to improve the channel estimation accuracy. For example, the coprime sampling concept has been incorporated into mmWave channel estimation to improve its performance in \cite{Caire2017TSP}.

In mmWave outdoor cellular systems, the user mobility may lead to huge Doppler effects which seriously deteriorates the performance of channel estimation \cite{Xiao2017JSAC}.
To the best of our knowledge, the joint Doppler and channel estimation for mmWave systems has not been extensively studied in literature.

In this paper, by fully utilizing the special structural characteristics of the sparse array, we first formulate an antenna selection based signal model for channel estimation, and then extend it to handle the Doppler shift. Based on this model, a joint AoA and Doppler estimation method is proposed by using the covariance fitting criterion, giving the path gains and channel estimates. We also provide an algorithmic implementation from the duality perspective to reduce computational complexity. Compared to other sparse methods, the proposed method does not require discretizing the angle space, and hence the grid mismatch effect can be completely eliminated. Numerical results are provided to verify the effectiveness of our proposed method.

Notations used in this paper are as follows. $\mathbb{C}$ denotes the set of complex numbers. $\bm{A}^T$ and $\bm{A}^H$ denote the transpose and conjugate transpose of $\bm{A}$, respectively. vec$(\bm{A})$ denotes the vectorization operator that stacks matrix $\bm{A}$ column by column. $\bm{A}\otimes\bm{B}$ is the Kronecker product of matrices $\bm{A}$ and $\bm{B}$. tr$(\bullet)$ is the trace operator. $\bm{I}_N$ denotes the identity matrix of size $N \times N$. $\|\bm{A}\|_F$ denotes the Frobenius norm of $\bm{A}$. For a positive semidefinite matrix $\bm{A}$, $\bm{A}\geq \bm{0}$ means that matrix $\bm{A}$ is positive semidefinite.

\section{Signal Model}
\label{sec signal model}

\begin{figure}[!t]
\centering
\includegraphics[width=2.35in]{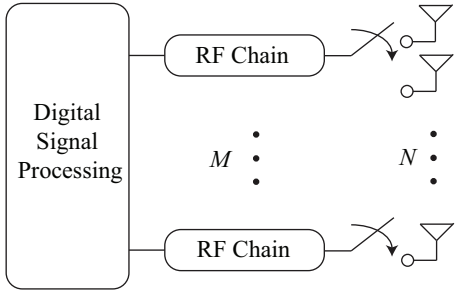}
\caption{mmWave architecture with switches.}
\label{Fig: mmWave architecture}
\end{figure}

\subsection{Signal Model with Static Users}

We consider a typical mmWave massive MIMO system with uniform linear arrays (ULAs). In particular, it is assumed that a BS equipped with an $N$-element ULA serves a single-antenna user. Let $K$ be the number of channel paths, $g_{k,t}$ be the time-varying channel coefficient of the $k$-th channel path at the $t$-th snapshot. Due to the poor scattering property of mmWave, the uplink CSI at time $t$ can be formulated as,
\equ{ \spl{ \bm{h}(t) &= \sum_{k=1}^K g_{k,t} \bm{a}(\theta_k) \\
&= \bm{A}\bm{g}(t)} \label{eq: h model} }
where $ \bm{a}(\theta_k) = [1, e^{j\pi\sin{\theta_k}},\cdots, e^{j(N-1)\pi\sin{\theta_k}}]^T $ is the steering vector associated with the AoA of the $k$-th channel path $\theta_k$, $\bm{A} = [\bm{a}(\theta_1),\cdots, \bm{a}(\theta_K)]$ is the manifold matrix of the array consisting of the steering vectors of the $K$ channel paths and $\bm{g}(t) = [g_{1,t},\cdots, g_{K,t}]^T $.\footnote{We assume that the AoAs are fixed during $L$ snapshots, which is reasonable and has been widely used in literature.} Assuming the training symbol is $s(t) $ with $|s(t)|=1$, the received signal multiplied by $s^*(t)$ at the BS can be represented as,
\equ{ \spl{
\bm{x}(t) &= s^*(t) \bm{h}(t) s(t) + \bm{n}(t)\\
&= \bm{h}(t) + \bm{n}(t), \label{eq: x model}
}}
where $\bm{n}(t)$ is the additive Gaussian white noise with zero mean and variance $\sigma\bm{I}$.

Unlike the traditional MIMO systems working at sub-6GHz band, the mmWave massive MIMO systems usually involve the hybrid analog-digital architecture. In this paper, we focus on the switch network based hybrid architecture, which is shown in Fig. \ref{Fig: mmWave architecture}. Let $M$ be the number of RF chains of the BS, where $M < N$. The switches connect the selected antennas to the RF chains. Obviously, the selection of the antennas, or the design of the selecting matrix, is a decisive factor for the estimation performance. For example, let $N = 7, M = 4$. If we select the antennas as $\bm{\Omega} = \{1,2,3,4\}$, which means that the antennas indexed as $1-4$ are active while others are inactive, we would not be able to obtain the true CSI with respect to the whole antennas \cite{PPal2011TSP_coprime}. So an important criterion of antenna selection is to estimate the CSI of the whole array by using the selected antennas only.

Thanks to the recent development of sparse array design in the area of array signal processing, the concepts of coprime array (CA), nested array (NA) and minimum redundancy array (MRA) can be employed to meet the criterion of antenna selection \cite{PPal2011TSP_coprime, zhouchengwei2017TVT, PPal2011TSP_Nested, zhouchengwei2017sensors}\cite{zhouchengwei2017IETcom}\cite{Zhouchengwei2017SJ}. In particular, an MRA with the selected antenna index set $\bm{\Omega} = \{1,2,5,7\}$ can be regarded as a virtual 7-element ULA from the covariance perspective. Then the entire CSI can be retrieved by using the MRA. In practice, since MRAs do not have simple closed-form expressions for the array geometry (i.e., given $N$ and $M$, it is in general difficult to derive the selected antenna index set $\bm{\Omega}$), and CAs have holes in the coarray, NAs become the best choice for antenna selection in analog precoding/combining design \cite{LiuChunlin2016TSP_SuperNested_I}.

\begin{figure}[!t]
\centering
\includegraphics[width=3.5in]{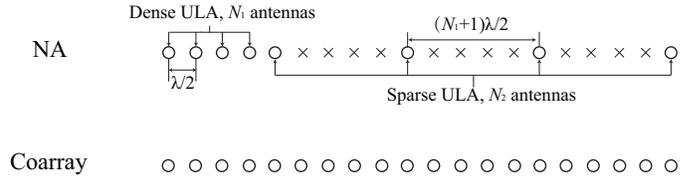}
\caption{NA and its coarray.}
\label{Fig: Nested array}
\end{figure}

A typical NA is composed of an $N_1$-element dense ULA and an $N_2$-element sparse ULA, which can be illustrated in Fig. \ref{Fig: Nested array}. It has been proved that the NA can be regarded as a ULA with $(N_1+1)N_2$ antennas from the coarray perspective \cite{PPal2011TSP_Nested}. The NA is suitable for the mmWave massive MIMO systems for three reasons: 1) the coarray of the NA is always a ULA with no hole; 2) the concept of NA can be conveniently incorporated into analog design by using switches; 3) the layout of the switch networks can be easily obtained given the number of antennas.

Formally, let $\bm{W}_{\bOmega} \in \mathbb{C}^{M\times N}$ be the selecting matrix (i.e., the analog combining matrix) such that the $m$-th row of $\bm{W}_{\bOmega}$ contains all $0$s but a single $1$ at the $\Omega_m$-th position, where $\bOmega = \{\Omega_1,\cdots,\Omega_M\}$ with $\Omega_1 = 1$ contains the selected antenna indices. Then by combining the received signal with $\bm{W}_{\bOmega}$, we have,
\equ{\label{eq:x model with Tau}
\spl{
  \bm{x}_{\bOmega}(t) &= \bm{W}_{\bOmega}\bm{x}(t) \\
  &= \bm{W}_{\bOmega}\bm{Ag}(t) + \bm{W}_{\bOmega}\bm{n}(t) \\
  &= \bm{A}_{\bOmega}\bm{g}(t) + \bm{n}_{\bOmega}(t),
}}
where $\bm{A}_{\bOmega} = \bm{W}_{\bOmega}\bm{A}$ and $\bm{n}_{\bOmega}(t) = \bm{W}_{\bOmega}\bm{n}(t)$.
Since $\bm{h}$ is associated with $\bm{A}$ and $\bm{g}(t)$, our target can be transformed to estimate $\bm{A}$ and $\bm{g}(t)$. Note that this model is quite similar to that in \cite{my2016TVT}, hence our previously proposed method \cite{my2016TVT} as well as other methods \cite{yangzai2015GLS} for DOA estimation can be easily extended to channel estimation.

\subsection{Signal Model with Fast-Moving Users}

When the users are moving, especially at a high speed, the Doppler effect should be taken into consideration. Formally, the received signal of the BS can be represented as,
\equ{\label{eq: x model with Doppler}
\bm{x}_{\bOmega}(t) = \sum_{k=1}^K g_{k,t} \bm{a}_{\bOmega}(\theta_k) e^{j2\pi f_k t} + \bm{n}_{\bOmega}(t),
}
where $f_k = \frac{T}{\lambda} v_k$ denotes the normalized Doppler frequency in which $T$, $\lambda$ and $v_k$ denote the sampling interval, wavelength and the radial velocity of the $k$-th sources, respectively. In this scenario, our target becomes to jointly estimate the AoAs $\theta_k$ and the Doppler shifts $f_k$ as well as the path gains $g_{k,t}$.

The channel model (\ref{eq: x model with Doppler}) is similar to the signal model for 2-dimensional (2-D) DOA estimation with single snapshot. Hence the method for 2-D DOA estimation can be directly extended to the channel estimation. However, since many 2-D DOA estimation methods require multiple snapshots, they are not suitable for channel estimation. In the following, we will propose a sparse channel estimation method for single-snapshot model (\ref{eq: x model with Doppler}) by using the covariance fitting criterion.

\section{Proposed Method}
\label{sec: proposed method}

Since the channel is fast time-varying, we only collect several snapshots for channel estimation (say, e.g., $3-5$). Hence it is reasonable to assume that $g_{k,t} = g_{k}$ in this paper. After collecting $L$ snapshots, we have,
\equ{ \label{eq: X model with Doppler}
\bm{X}_{\bOmega} = \sum_{k=1}^K g_k \bm{a}_{\bOmega}(\theta_k) \bm{v}^T(f_k) + \bm{N}_{\bOmega},
}
where $\bm{v}(f_k) = [e^{j2\pi f_k},\cdots, e^{j2\pi L f_k}]^T$.
Vectorizing (\ref{eq: X model with Doppler}) we have,
\equ{\spl{
\bm{y}_{\bOmega} &= \text{vec}(\bm{X}_{\bOmega}) \\
&= \sum_{k=1}^K g_k \underbrace{\bm{v}(f_k) \otimes \bm{a}_{\bOmega}(\theta_k)}_{\bm{b}_{\bOmega k } } + \bm{e}_{\bOmega} \\
&= \sum_{k=1}^K g_k (\underbrace{\bm{I}_L\otimes \bm{W}_{\bOmega}}_{\bgo})(\underbrace{\bm{v}(f_k)\otimes \bm{a}(\theta_k)}_{\bm{b}_k}) + \bm{e}_{\bOmega} \\
&= \bgo \sum_{k=1}^K g_k \bm{b}_k +\bm{e}_{\bOmega} \\
&= \bgo \bm{y} +\bm{e}_{\bOmega}
}}
where $\bm{b}_{\bOmega k } = \bgo \bm{b}_k$, $\bgo \in \bC^{ML \times NL}$ can be regarded as the two-dimensional selecting matrix, $\bm{y} = \sum_{k=1}^K g_k \bm{b}_k$ denotes the full received data vector of the entire $N$-element ULA and $\bm{e}_{\bOmega} = \text{vec}(\bm{N}_{\bOmega}) = \bgo\text{vec}(\bm{N})$.
The covariance matrix of $\bm{y}_{\bOmega}$  can then be obtained as,
\equ{\spl{
\bm{R}_{\bOmega} &= E[\bm{y}_{\bOmega} \bm{y}^H_{\bOmega}] \\
&= \sum_{k=1}^K p_k \bm{b}_{\bOmega k} \bm{b}^H_{\bOmega k} + E[\bm{e}_{\bOmega}\bm{e}_{\bOmega}^H] \\
&= \bgo \sum_{k=1}^K p_k \bm{b}_k \bm{b}_k^H \bgo^H + \sigma \bgo\bgo^H \\
&= \bgo \bT \bgo^H \\
&= \bT_{\bOmega} ,
}}
where $p_k = E[|g_k|^2]$, $\bT \in \bC^{NL \times NL}$ is a positive semidefinite (PSD) Toeplitz-block-Toeplitz matrix, $\bT_{\bOmega}$ can be regarded as a submatrix of $\bT$. It can be seen that, the variables $\theta_k$ and $f_k$ are encoded in $\bT$, hence an accurate recovery of the covariance matrix $\bT$ is helpful if $\theta_k$ and $f_k$ can be correctly retrieved from $\bT$. Fortunately, we have the following multiple Vandermonde decomposition theorem to ensure a successful reconstruction of $\theta_k$ and $f_k$.
\begin{theorem}[\cite{YANG2016MD}] \label{thm:vanderdec}
  Given a PSD Toeplitz-block-Toeplitz matrix $\mathbb{T} \in \mathbb{C}^{NL \times NL}$ with $K\leq\min\{N, L\}$ being the rank, then $\bT$ satisfies the following decomposition,
  \begin{equation}\label{eq decomposition}
    \bT = \sum_{k=1}^K p_k \big( \bm{v}(f_k)\bm{v}^H(f_k) \big) \otimes \big( \bm{a}(\theta_k)\bm{a}^H(\theta_k) \big) + \sigma \bm{I},
  \end{equation}
  where $\sigma$ denotes the smallest eigenvalue of $\bT$, $p_k>0$, and moreover this decomposition is unique.
\end{theorem}
It is pointed out in \cite{YANG2016MD} that, the decomposition of $\bT$ can be guaranteed with a high probability if $K< NL - \max\{N,L\}$, which will be verified in simulations.
Hence, our remaining work is to accurately reconstruct the covariance matrix $\bT$. Especially, it has a special structure of Toeplitz-block-Toeplitz, which could be used to improve the estimation performance.\footnote{Interested readers are referred to \cite{YANG2016MD} for more details about the properties of Toeplitz-block-Toeplitz structure.}

Inspired by the method in \cite{yangzai2015GLS}, we propose the following covariance fitting criterion,
\equ{\spl{
h = &\norm{\bm{R}_{\bOmega}^{-\frac{1}{2}} (\bm{y}_{\bOmega} \bm{y}_{\bOmega}^H - \bm{R}_{\bOmega}) }_F^2 \\
= & \norm{\bm{y}_{\bOmega}}_2^2 \bm{y}_{\bOmega}^H \bm{R}_{\bOmega}^{-1} \bm{y}_{\bOmega} + \text{tr}(\bm{R}_{\bOmega}) - 2\norm{\bm{y}_{\bOmega}}_2^2.
}}
Minimizing $h$ will result in the following semidefinite programming (SDP),
\equ{ \label{eq: SDP} \spl{
&\min h \\
= &\min_{\bT} \norm{\bm{y}_{\bOmega}}_2^2 \bm{y}_{\bOmega}^H \bT_{\bOmega}^{-1} \bm{y}_{\bOmega} + \text{tr}(\bT_{\bOmega}) \\
= &\min_{z,\bT} \norm{\bm{y}_{\bOmega}}_2^2 z + \text{tr}(\bT_{\bOmega}),\ \text{s.t.}\ z\geq \bm{y}_{\bOmega}^H \bT_{\bOmega}^{-1} \bm{y}_{\bOmega} \\
= &\min_{z,\bT} \norm{\bm{y}_{\bOmega}}_2^2 z + \text{tr}(\bT_{\bOmega}),\ \text{s.t.}\ \bma{ z & \bm{y}_{\bOmega}^H\\ \bm{y}_{\bOmega} & \bT_{\bOmega} } \geq \bm{0}.
}}
Once the SDP is solved, the parameters of interest $\theta_k$ and $f_k$ can be retrieved according to Theorem \ref{thm:vanderdec}. In order to determine the path gains of the channel, we substitute the estimates $\hat{\theta}_k$ and $\hat{f}_k$ into (\ref{eq: X model with Doppler}), leading to the least squares solution of $\hat{\bm{g}} = [\hat{g}_1,\cdots,\hat{g}_K]$.

Finally, given the estimated AoAs and path gains, the full CSI of the BS can be obtained as,
\equ{\label{eq: h final}
\hat{\bm{h}} = \sum_{k=1}^K \hat{g}_k \bm{a}(\hat{\theta}_k).
}
The proposed method for joint Doppler and channel estimation can be summerized in Algorithm \ref{alg}.
\begin{algorithm}[h]
\caption{Proposed method}
\label{alg}
\begin{algorithmic}
\REQUIRE received data $\bm{X}_{\bm{\Omega}}$.
\STATE  step 1: Obtain the covariance matrix $\hat{\bT}$ by solving the problem (\ref{eq: SDP});
\STATE  step 2: Obtain $\hat{\theta}_k$ and $\hat{f}_k$  by applying Theorem \ref{thm:vanderdec} to $\hat{\bT}$;
\STATE  step 3: Obtain the path gains $\hat{\bm{g}}$ from (\ref{eq: X model with Doppler}),
\STATE  step 4: Obtain the channel estimate $\hat{\bm{h}}$ from (\ref{eq: h final}),
\ENSURE $\hat{\bm{h}}$ and $\hat{f}$.
\end{algorithmic}
\end{algorithm}

\begin{remark}
  Although our proposed method is closely related to the method proposed in \cite{liuchunlin2015ICASSP_JADE}, yet the superiority of our method is obvious. The method JADE in \cite{liuchunlin2015ICASSP_JADE} employs the coprime sampling concept, while our proposed signal model can be applied to all kinds of sparse arrays, e.g., CA, NA, MRA and arbitrary linear arrays. Moreover, JADE uses the spatial smoothing procedure to deal with the coherent sources, which may reduce the degrees of freedom. In contrast, our method can be directly used for coherent sources
\end{remark}
\begin{remark}
  Our method can be also related to the method in \cite{doppler_coprime_atomic2016RADAR}, which involves the atomic norm minimization (ANM) method and coprime sampling concept to jointly estimate AoA and Doppler. However, the ANM method has to know the exact value of the noise power, which is usually unknown in practice. Our proposed method does not require the knowledge of the noise level and hence is more practically attractive. Moreover, since the SDP to be solved in (\ref{eq: SDP}) has a smaller model dimension than that in \cite{doppler_coprime_atomic2016RADAR}, our method is more computationally efficient, which will be illustrated via simulations.
\end{remark}
\begin{remark}
  Our method can be easily extended to the scenario of planar arrays. The only difference is that $\bm{a}_{\bOmega}(\theta_k)$ has to be replaced by the one with respect to the 2-D vectorized steering vector \cite{mySJ2017FGML}, resulting a three-level Toeplitz covariance matrix. We can also formulate a similar reconstruction model as (\ref{eq: SDP}) and solve it using CVX. Finally the azimuth and elevation AoAs and Doppler shifts can be jointly retrieved by using multiple Vandermonde decomposition theorem \cite{YANG2016MD}. The details will be investigated in our future study.
\end{remark}

\section{Algorithmic Implementation From Duality}
\label{sec: duality}

The SDP (\ref{eq: SDP}) can be efficiently solved by using CVX. Alternatively, we also observe that solving the dual problem of (\ref{eq: SDP}) is more computationally efficient than solving the original one. In the following, we will derive the dual problem of (\ref{eq: SDP}).

Firstly, let $\bm{Q} = \bma{u & \bm{w}^H \\ \bm{w} & \bm{V}}\geq \bm{0}$ be the Lagrangian multiplier of the constraint of (\ref{eq: SDP}). Then the Lagrangian of (\ref{eq: SDP}) can be written as,
\equ{\spl{
&\mathcal{L}(\bT,z,\bm{Q}) \\
= &  \|\bm{y}_{\bOmega}\|_2^2 z + \text{tr}(\bT_{\bOmega}) - \text{tr}\left\{ \bma{u & \bm{w}^H \\ \bm{w} & \bm{V}} \bma{z & \bm{y}_{\bOmega}^H\\ \bm{y}_{\bOmega} & \bT_{\bOmega} } \right\} \\
= &  \|\bm{y}_{\bOmega}\|_2^2 z + \text{tr}(uz+\bm{w}^H \bm{y}_{\bOmega} + \bm{w}\bm{y}_{\bOmega}^H + \bm{V}\bT_{\bOmega}) \\
= &  (\|\bm{y}_{\bOmega}\|_2^2 - u)z + \text{tr}[(\bm{I}-\bm{V})\bT_{\bOmega}] - 2\Re\{\text{tr}(\bm{w}^H\bm{y}_{\bOmega})\} \\
= &  (\|\bm{y}_{\bOmega}\|_2^2 - u)z + \text{tr}[\bgo^H(\bm{I}-\bm{V})\bgo\bT] - 2\Re\{\text{tr}(\bm{w}^H\bm{y}_{\bOmega})\} .\\
}}
Thus, the Lagrange dual to (\ref{eq: SDP}) can be formulated as,
\equ{\spl{
\mathcal{G} = & \max_{\bm{Q}} \min_{\bT,z} \mathcal{L}(\bT,z,\bm{Q}) \\
= & \min_{\bm{Q}} 2\Re\{\text{tr}(\bm{w}^H\bm{y}_{\bOmega})\}\ \text{s.t.} \cas{\bm{Q} \geq \bm{0} \\
u = \|\bm{y}_{\bOmega}\|_2^2 \\ \bT^*(\bgo^H (\bm{I} - \bm{V}) \bgo) = \bm{0},}
}}
which can also be solved by using CVX. Since a strong duality holds, the dual variable of $\bm{Q}$ is equivalent to $\bT_{\bOmega}$. Finally, noting that there is one-to-one correspondence between $\bT_{\bOmega}$ and $\bT$ \cite{yangzai2015GLS}, the optimal solution of problem (\ref{eq: SDP}) can be obtained from $\bT_{\bOmega}$.

\section{Numerical Results}
\label{sec: numerical results}

\begin{figure}[!t]
\centering
\includegraphics[width=2.8in]{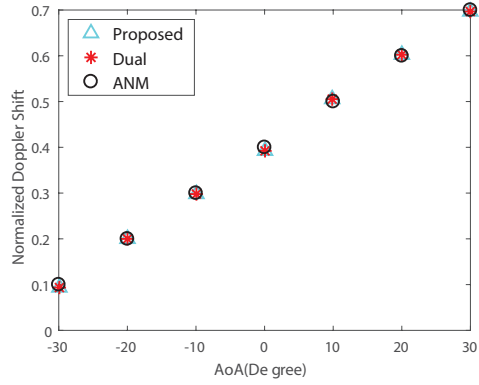}
\caption{Estimation performance of the proposed method with $ N = 32$, $M = 11$, $L = 5$ and $K = 7$.}
\label{Fig: one trial}
\end{figure}

\begin{figure}[!t]
\centering
\includegraphics[width=2.8in]{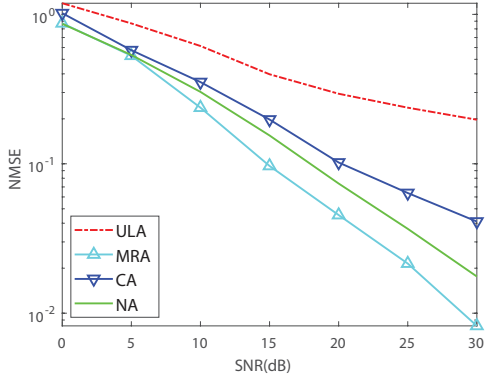}
\caption{NMSE comparison of different array structures with $M = 6$, $L = 5$ and $K = 3$.}
\label{Fig: different array}
\end{figure}

\begin{figure}
\centering
\subfigure[NMSE comparison]{
\begin{minipage}[b]{0.40\textwidth}
\includegraphics[width=1\textwidth]{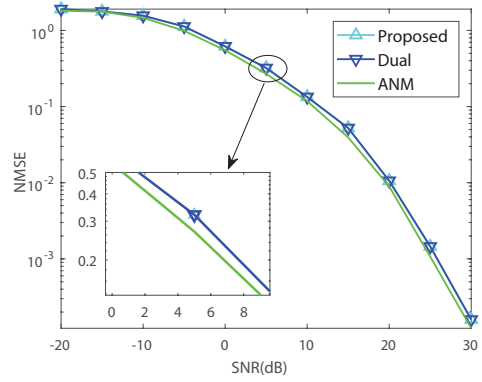} \\
\end{minipage}
}
\subfigure[Running time comparison]{
\begin{minipage}[b]{0.40\textwidth}
\includegraphics[width=1\textwidth]{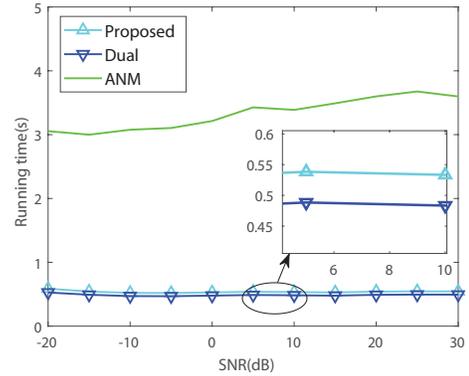} \\
\end{minipage}
}
\caption{NMSE and running time comparisons of our proposed method and ANM with $ N = 32$, $M = 11$, $L = 3$ and $K = 3$.}
\label{Fig: NMSE and running time}
\end{figure}
%

In this section, we will evaluate the performance of our proposed method as well as its dual realization. We choose the ANM method in \cite{doppler_coprime_atomic2016RADAR} for performance comparison. The regularization parameter in ANM is determined according to the noise power \cite{AST2013Bhaskar}.

First we provide a simple example to show the estimation performance of our method in terms of AoA and Doppler. We consider a typical mmWave massive MIMO system, where the BS uses a $32$-element ULA and $11$ RF chains to serve a single-antenna user. The antenna selection strategy at BS is chosen as $\bOmega = \{1,2,3,4,5,6,7,8,16,24,32\}$ according to the NA structure. The number of channel paths is set to $7$. The AoAs and the normalized Doppler shifts are randomly chosen from the ranges of $[-30^{\circ},30^{\circ}]$ and $[0.1, 0.7]$, respectively. Considering that the channel is fast time-varying, we only collect $5$ snapshots for joint AoA and Doppler estimation, which is smaller than the number of paths. The estimates of both AoAs and Doppler shifts are provided in Fig. \ref{Fig: one trial}, from which it can be seen that all the (AoA, Doppler) pairs can be correctly estimated and the dual realization yields the same solution to the original problem.

We then compare the estimation performance of different array structures. In particular, we fix the number of RF chains as $11$ and construct four antenna selection strategies: ULA, MRA, CA and NA. The normalized mean square errors (NMSEs) of channel estimates with respect to these strategies are illustrated in Fig. \ref{Fig: different array} with the SNR varying from $0$dB to $30$dB. It can be seen that due to the largest array aperture, the MRA enjoys the best estimation performance. However, the MRA does not have an exact formulation for the array structure. Moreover, the formulation may not exist given a certain number of antennas. In contrast, NA shows suboptimal performance and its structure can be easily obtained. CA and ULA provide inferior performance since their continuous coarray is shorter than the other two strategies.

We now compare our method with respect to ANM. Let $ N = 32$, $M = 11$, $L = 3$ and $K = 3$. The AoAs and the normalized Doppler shift are randomly chosen from the range of $[-30^{\circ},30^{\circ}]$ and $[0.1,0.7]$, respectively. We evaluate the NMSE of the channel estimates and running time with the SNR varying from $-20$dB to $30$dB and show the simulation results in Fig. \ref{Fig: NMSE and running time}.

From Fig. \ref{Fig: NMSE and running time}(a) we can see that, the performance of the proposed method is quite similar to that of ANM and the NMSE of the compared methods decreases as the SNR grows. The reason why ANM is slightly better than our method is that, ANM employs the accurate noise power which is unrealistic. Also, from Fig. \ref{Fig: NMSE and running time}(a), it can be observed that solving the dual problem yields the same estimation performance as solving the original one. The running time result shown in Fig. \ref{Fig: NMSE and running time}(b) reveals that, our method enjoys a much higher computational efficiency than ANM in the whole SNR region. This is because the semidefinite constraint of ANM is of size $(N+1)\times (N+1)$. As a result, the SDP for recovering the Toeplitz-block-Toeplitz structure has time complexity $O(N^{3.5}\log{1/\epsilon})$, where $\epsilon$ denotes the desired recovery precision \cite{Tianzhi2017ICC_truncated}. In contrast, the semidefinite constraint of our method is of size $(M+1)\times (M+1)$ and hence the time complexity is $O(M^{3.5}\log{1/\epsilon})$, which is much smaller than that of ANM. Finally, from Fig. \ref{Fig: NMSE and running time}(b) we can also observe that, more than $10\%$ running time can be saved if we solve the dual problem rather than the original one.

\begin{figure}[!t]
\centering
\includegraphics[width=2.8in]{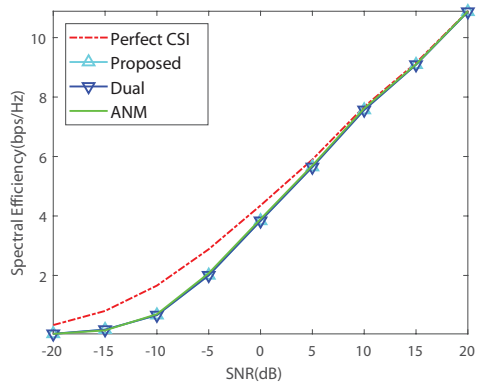}
\caption{Spectral efficiency comparison of the proposed method and ANM with $ N = 32$, $M = 11$, $L = 3$ and $K = 3$.}
\label{Fig: SE SNR}
\end{figure}

Finally, we compare the spectral efficiency of these methods and show the results in Fig. \ref{Fig: SE SNR}. The case with ideal CSI is adopted as the upper bound for performance comparison. It can be observed that, the three methods show very similar performance and are able to coincide with the upper bound when the SNR is larger than $10$dB. Then we can conclude that our method is superior to ANM due to the fact that our method is much more time-saving and without requiring the knowledge of noise.

\section{Conclusion}
\label{sec: conclusion}

In this paper, the channel estimation for mmWave communications in fast varying environment was considered and boiled down to joint AoA and Doppler estimation. According to the NA structure, a low-complexity switch network is provided for analog architecture design. The NA structure can improve the channel estimation performance, as verified by simulations. A joint AoA and Doppler estimation method was proposed based on the covariance fitting criterion. The proposed method does not discretize the angle space and hence can totally eliminate the model mismatch effect caused by discretization. Moreover, our method does not need the noise power as \emph{a priori} and hence is more attractive in practical applications.

\appendices

\section*{Acknowledgment}

This work was supported by the National Natural Science Foundation of China under grant No. 61471205 and No. 61771256, the Key International Cooperation Research Project under Grant No. 61720106003.

\bibliographystyle{IEEEtran}
\bibliography{reference}

\end{document}